\documentclass[12pt]{iopart}

\usepackage{graphicx}
\usepackage{amsfonts}

\begin{document}
\title{The Impacts of Subsidy Policies on Vaccination
Decisions in Contact Networks}

\author{Hai-Feng Zhang$^{1,2}$, Zhi-Xi Wu$^{3}$, Xiao-Ke Xu$^{4}$, Michael Small$^{5}$, Bing-Hong Wang$^{6}$}

\address{$^{1}$School of Mathematical Science, Anhui University,
Hefei 230601, China}
\address{$^{2}$ Department of Electronic and Information Engineering, Hong
Kong Polytechnic University, Hung Hom, Kowloon, Hong Kong, China}
\address{$^{3}$ Institute of Computational Physics and Complex Systems,
Lanzhou University, Lanzhou, China}
\address{$^{4}$ College of Information and Communication Engineering, Dalian Nationalities University; Dalian 116605; China
}
\address{$^{5}$ School of Mathematics and Statistics, University of Western
Australia, Crawley, Australia}
\address{$^{6}$ Department
of Modern Physics, University of Science and Technology of China,
Hefei, China}
\ead{haifeng3@mail.ustc.edu.cn (HFZ);
michael.small@uwa.edu.au (MS)}
\begin{abstract}
Often, vaccination programs are carried out based on self-interest
rather than being mandatory. Owing to the perceptions about
risks associated with vaccines and the `herd immunity' effect, it may
provide suboptimal vaccination coverage for the population as a whole. In this
case, some subsidy policies may be offered by the government to
promote vaccination coverage. But, not all subsidy policies are
effective in controlling the transmission of infectious diseases.
We address the question of which subsidy policy is best, and how
to appropriately distribute the limited subsidies to maximize
vaccine coverage. To answer these questions, we establish a model
based on evolutionary game theory, where individuals try to
maximize their personal payoffs when considering the voluntary
vaccination mechanism. Our model shows that voluntary vaccination
alone is insufficient to control an epidemic. Hence, two subsidy
policies are systematically studied: (1) in the free subsidy
policy the total amount of subsidies is distributed to some
individuals and all the donees may vaccinate at no cost, and (2)
in the part-offset subsidy policy each vaccinated person is offset
by a certain proportion of the vaccination cost. Simulations
suggest that, since the part-offset subsidy policy can encourage
more individuals to be vaccinated, the performance of this policy
is significantly better than that of the free subsidy policy.
Meanwhile, for the free subsidy policy, if the individuals with
more neighbors are freely vaccinated with high priority rather
than in a random manner, we find that the efficiency of this
subsidy policy is remarkably improved. 
In addition, for the part-offset subsidy policy, we observe that
too low or too high proportion of subsidy is not good for
group-optimal vaccination given that the total amount of subsidies
is restricted. Instead, moderate proportion of allowances can
achieve the maximum of the social benefits.
\end{abstract}

 \maketitle
\tableofcontents
\section{Introduction} \label{sec:intro}
Preemptive vaccination is the fundamental method for preventing
transmission of infectious diseases as well as reducing morbidity
and mortality. Many proposed strategies, however, build upon a major
premise that the vaccination or immunization is compulsory and have
not considered the willingness or desire of individuals, such as target immunization~\cite{pastor2003epidemics}, ring vaccination~\cite{muller2000ring} and acquaintance immunization~\cite{cohen2003efficient}
Practically, the immunization of individuals is more of a
voluntary behavior. If individuals are permitted to choose
whether to vaccinate according to self-interest, individual
behavioral responses play a key role in determining the success of
the vaccination program and its consequences for disease
control~\cite{Bauch03,Bauch04,Shim11,Bauch09,Reluga11}.
Meanwhile, the likelihood of choosing vaccination is affected by
many factors, including: perceived risk of infection; cost of
vaccination; and, the vaccination behavior of other individuals.
Many previous studies, based on the imitation
rule~\cite{Bauch05,Bauch10,Fu1,Fu2}, the payoffs maximization
rule~\cite{Bauch11,Galvani07}, or the minority
game rule to investigate the vaccination
dynamics~\cite{Blower1,Blower2,Blower3}, have concluded that it
would be difficult or even impossible to eradicate a
vaccine-preventable disease under a voluntary vaccination
mechanism without additional incentives. The reason is that, with
increasing coverage of vaccination in the community, the
individuals who are unvaccinated are less likely to become
infected, and therefore, they have less incentive to take a
vaccine, and then due to the existence of free-riders, the actual
coverage of vaccination often cannot meet the social
demand~\cite{Bauch03,Bauch04}.

Though many authors have addressed the key role of human
behavioral responses on vaccination dynamics~\cite{funk10,meloni2011modeling,gross2006epidemic,meloni2009traffic}, to the
best of our knowledge, the influences of different subsidy
policies on vaccination decisions have not been thoroughly
investigated. In many cases, on account of the gap between the
actual coverage of vaccination and the social demand, some
incentives, such as subsidy policies should be offered by the
government to induce more people to take vaccination. Faced with
various choices, which subsidy policy is the best one? How can the
efficiency of the subsidy policy be maximized with a restricted
amount of available subsidy? In this work we move towards
answering these questions.

Previous studies of the interplay between diseases dynamics and
individuals' vaccination behaviors have often assumed that
diseases are transmitted in either homogeneously mixed populations
or random networks. A recent important advancement, is the
application of network theory to study epidemic dynamics
~\cite{network1,liu2010heat,network2}. Network-based models have been
shown to be able to accurately reflect the heterogeneities in the
number of contacts of individuals~\cite{masuda2010effects,holme2004efficient,network}.

Motivated by these facts, in the present work we first develop a
network-based vaccination model in which individuals update their
vaccination strategies by balancing the advantages and
disadvantages of vaccination. And we find
that, owing to the `herd immunity' effect in voluntary vaccination
mechanism, it is difficult to eradicate the diseases. Hence, two
types of subsidy policies are considered: In the first case
(labeled free subsidy policy), a certain fraction of individuals
are freely vaccinated according to the total amount of subsidies
and the cost of vaccination; In the second case (labeled
part-offset subsidy policy), a certain proportion of allowances
are subsidized to all vaccinated persons. We show that the free
subsidy policy is not helpful for improving the probabilities of
vaccination among non-donees, but actually reduce it. In contrast,
the part-offset subsidy policy can encourage more individuals to
take vaccination, leading to the better performance of this policy
when the same amount of subsidies is supplied. Furthermore, we
verify that the efficiency of the free subsidy policy can be
significantly improved when the donees are chosen in a descending
degree order rather than in a random way. In addition, for the
part-offset subsidy policy, we find that there exists optimal
proportion of allowance inducing the group-optimal phenomenon with
the fixed amount of subsidies.


\section{Vaccination rule and subsidy policies}\label{sec:model}
\subsection{Vaccination update rule}\label{sec:vaccination}
We model the interplay between vaccination behavior and the
transmission dynamics on contact networks, where the nodes
represent individuals and the edges represent the contacts between
them, along which an infection can be transmitted. For simplicity,
the networks remain static in our study. The number of
neighbors of a node is called its degree, and the distribution of
these values is called the degree distribution~\cite{Bauch12a}.

In our model, individuals need to decide whether to vaccinate or
not before implementing the vaccination programs. Usually, the
formation of the vaccination opinion is affected by many factors,
such as: the side-effects of the vaccines; fear~\cite{ff,wang2011effects}; availability of
information concerning the epidemic~\cite{do1,zhang2012modeling}; and, the opinions of
other persons, to name just a few~\cite{FHC09,FHC11}. Here, we assume that individuals update
their vaccination decisions by maximizing their perceived benefits
(or in other words, by reducing their potential loss). The
perceived benefits of infection and the perceived benefits of
vaccination are given as~\cite{haifeng}
\begin{eqnarray}
  P(X_{i}=1) &=&-C_V; \label{eq1} \\
  P(X_{i}=0) &=&-C_I\lambda_i\label{eq2}.
\end{eqnarray}

Here $X_i=1$ or 0 denotes, respectively, that the individual $i$
accepts the vaccination or not. The parameters $C_V$ and $C_I$ are the
cost of vaccination and infection, and $\lambda_i$ is the
perceived probability of infection. Let $c=C_V/C_I$ be the
relative cost of vaccination, then Eqs.~(\ref{eq1}) and
(\ref{eq2}) can be rewritten as
\begin{eqnarray}
  P(X_{i}=1) &=&-c; \label{eq3} \\
  P(X_{i}=0) &=&-\lambda_i\label{eq4}.
\end{eqnarray}

A more sophisticated definition of the perceived probability of
infection $\lambda_i$ by combing a more diverse range of factors may offer a more precise
 model of reality, but would also increases the complexity of the
model. Here we define $\lambda_i$ for simplicity as a function of
transmission rate $\beta$ and the number of non-vaccinated
neighbors $k_{nv}^i$~\cite{Bauch12a},
\begin{equation}\label{eq5}
\lambda_i=1-(1-\beta)^{k_{nv}^i}.
\end{equation}
 Eq.~(\ref{eq5}) indicates that the greater the number of non-vaccinated
individuals within the neighborhood, the higher the probability of being
infected.

Following common practice, we assume the individuals are of finite
rationality. That is, the individuals prefer to choose strategies
with higher payoffs (payoff maximization rule), while it is
possible, but unlikely, that the individuals can make worse
decisions on the vaccination strategy. An usual way to achieve the
above purpose is to incorporate stochastic element into the model:
the probability of vaccination for the individual $i$ is given as~\cite{perc2012sustainable,szolnoki2011group}:
\begin{equation}\label{eq6}
p=\frac{1}{1+exp(-\alpha(P(X_i=1)-P(X_i=0)))},
\end{equation}
where $\alpha$ represents the degree that the individuals respond
to the difference of payoffs, the larger $\alpha$ the more
sensitive of the individuals. Our main results are obtained with
$\alpha=5$. We have verified that our results are robust to the moderate values of $\alpha$
(i.e. $1\leq\alpha\leq30$).

\subsection{Model description}\label{sec: model}
For each run of simulations, the following processes are
implemented sequentially~\cite{Bauch12a,opinion1,opinion3}:\\

(1)~ Construct a contact network. Empirical studies have shown that
the degree distributions of many social, biological, and
technological networks are of power-law forms rather than
Poissonian form. Here we use the classical scale-free
network---Barab\'asi-Albert network~\cite{BA} (labeled BA network)
with average degree $\langle k\rangle=6$ and $N=2000$ as the
underlying structure of contact network (We have also checked our results on other types of networks, such as the typical Poissonian
network---Erd\"os-R\'enyi random network~\cite{ER}, and we found that our results are robust to the structure of contact networks.).

(2)~ Formation of vaccination opinion. Each individual is
equally assigned a vaccination decision (i.e., vaccinate or not
vaccinate) such that the initial vaccination coverage is about
50\%. Then the individuals update their vaccination decisions with
the probability in Eq.~(\ref{eq6}). Such update rule is carried
out in a parallel way until a steady state is reached (1000
iterations for our model). Finally, each individual makes a
vaccination choice and the vaccinated individuals are immune to
infection. After that, we start the epidemic process.\\

(3)~ Epidemic process. The standard SIR
(susceptible--infected--removed) epidemic model is run on the
network generated in step (1) with the final vaccination decisions
obtained in step (2). Initially, $5$ individuals are randomly
chosen from non-vaccinated individuals as the seeds of infection
to begin the epidemic process, and others are either susceptible
or vaccinated individuals. At each time step, every susceptible
individual $i$ is infected with probability
$\Lambda_i=1-(1-\beta)^{k_{inf}^i}$, where $k_{inf}^i$ is the
number of infected neighbors of the individual $i$. An infected
individual recovers and becomes immune with probability $g$ per time
step. We run the model until no infection exists in the system.
Here, we calibrate the value of disease transmission probability
to ensure the final epidemic size to be equal to 90\% in the case
of non-vaccination in the population. We can achieve this by
fixing $g=0.25$, $\beta=0.18$ for BA network.

The equilibrium data are obtained by
averaging over 20 realizations for the epidemic processes, and 100
independent simulations for each process.

\section{Effects of subsidy policies}\label{sec:results}
\begin{figure}
\begin{center}
\includegraphics[width=4in]{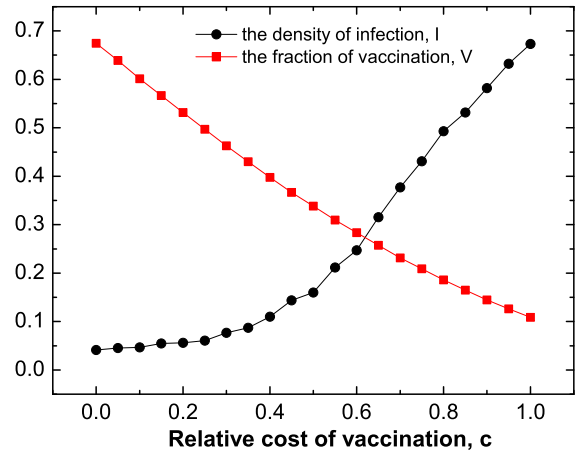}
\caption{
 Without any incentive, the density of
infection and the fraction of vaccination versus the relative cost of
vaccination $c$.  The fraction of vaccination decreases with the value of $c$, and which leads to the prevalence of disease.
}
\label{fig1}
\end{center}
\end{figure}

Firstly, taking no account of the government subsidy, the effect of
the relative cost of vaccination $c$ on the density of infection
($I$) and the fraction of vaccination ($V$) are presented in figure~\ref{fig1}. As
expected, the fraction of vaccination decreases with the value of
$c$, leading to the rapid increase of the density of infection.
Figure~\ref{fig1} also demonstrates that the epidemic can not be
eradicated if individuals make vaccination decisions
according to self-interest.

\begin{figure}
\begin{center}
\includegraphics[width=6in]{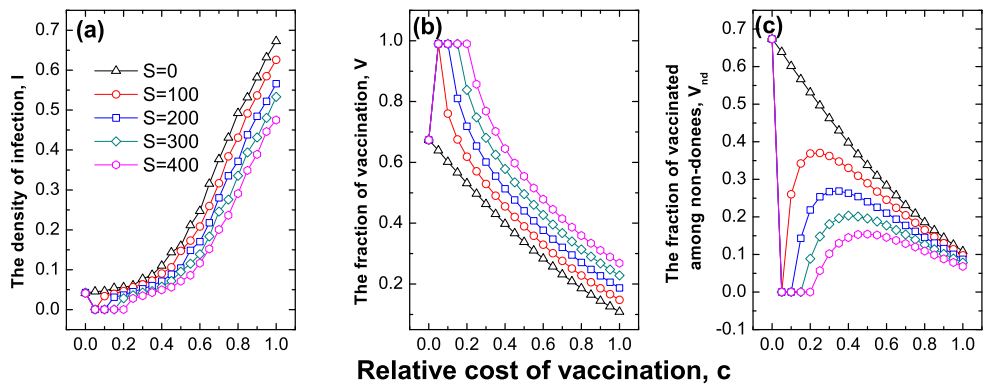}
\caption{
 The effect of free subsidy policy amount
of subsidies $S$ on the vaccination behaviors and the epidemic
size: (a) the density of infection; (b) the fraction of
vaccination and (c) the fraction of vaccinated among non-donees as
functions of the relative cost of vaccination $c$ for different
values of $S$. For simplicity, we select $S=100$, 200 or other values
in our studies, but the value of $S$ is not always fixed. Taking
$S=200$ as an example, when $c=0$, the amount of subsidy will be zero;
when $c$ is very small (e.g. $c=0.05$) then $S/c=4000>N=2000$, so all
the individuals could be freely vaccinated, and in such case the
actual amount of subsidy
is $S=cN=100$; when $S/c<N$, then we have $S=200$ and we still label $S=200$.
}
\label{fig2}
\end{center}
\end{figure}

In what follows, two different subsidy policies are compared to
emphasize the roles of subsidy policies on the epidemic control.
For the free subsidy policy, the total amount of subsidies (labeled $S$)
is distributed to some people and all the donees are vaccinated free of charge.
For example, if the total amount of subsidy $S=100$
and $c=0.5$, then 200 persons would be vaccinated at no personal cost. One
counter-intuitive feature is displayed in figure~\ref{fig2}: the fraction of
vaccination ($V$) is not improved remarkably even though the values of $S$
increase from 0 (without subsidy policy) to 400 (see
figure~\ref{fig2}(b)). As a result, such free subsidy policy has
limited effectiveness on reducing the density of infection ($I$)
(see figure~\ref{fig2}(a)).

To better explain this disadvantageous situation, the
rate of vaccination among non-donees ($V_{nd}$) is presented in
figure~\ref{fig2}(c). One can observe that, $V_{nd}$ does not
increase with the total amount of subsidies $S$, but rather
decreases with it. According to Eq.~(\ref{eq6}), the perceived risk
of infection by those non-donees will decrease when more
individuals are freely vaccinated. As a result, those non-donees prefer
to take risks after balancing the advantages and disadvantages of
vaccination. (For fixed value of $S$, we have $S/c\geq N$ when $c$
is very small. According to the rule of our model, all the
individuals will be freely vaccinated (see figure~\ref{fig2}(b)),
and this will lead to $V=0$ in figure~\ref{fig2}(c)).

\begin{figure}
\begin{center}
\includegraphics[width=6in]{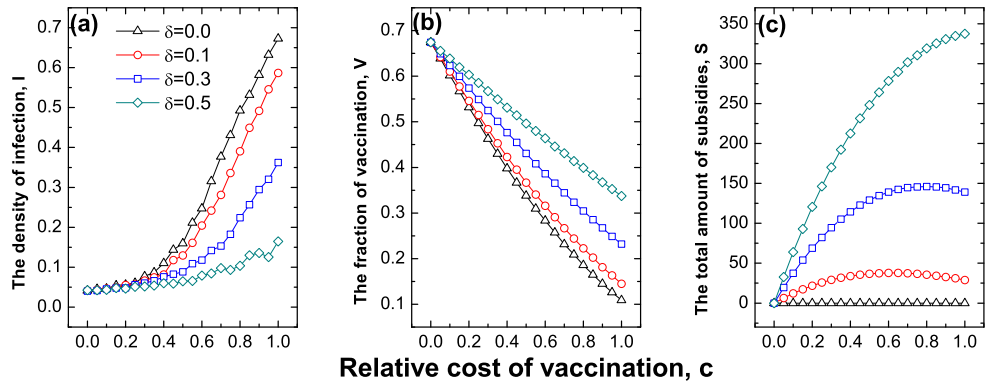}
\caption{
 The effect of part-offset subsidy policy
on the vaccination behaviors and the epidemic size: (a) the
density of infection; (b) the fraction of vaccination and (c) the
total amount of subsidies as functions of the relative cost of
vaccination $c$ for different proportion $\delta$.
}
\label{fig3}
\end{center}
\end{figure}

For the part-offset subsidy policy, the proportion $\delta c$ of
subsidy cost is offset by the government for each vaccinated individual.
That is, each vaccinated individual only needs to pay for
$(1-\delta)c$ when the cost of vaccination is $c$. In this case,
the effect of the part-offset subsidy policy on controlling the
spreading of epidemic is presented in figure~\ref{fig3}, where the
fraction of vaccination increases with the value of $\delta$ (see figure~\ref{fig3}(b)). And this leads to a sharp decline in the density of
infection (see figure~\ref{fig3}(a)). For example, when $c=1.0$, the
density of infection is reduced from about 0.65 to 0.15 with $\delta$ increasing from 0 to 0.5. In particular, as shown
in figure~\ref{fig3}(c), the total amount of subsidy is rather low on the
premise of the efficiency of subsidy. For instance, even 50\% of
cost is subsidized, the total amount of subsidy $S$ will be smaller
than 350. Comparing $S=400$ in figure~\ref{fig2}(a) with $\delta=0.5$
in figure~\ref{fig3}(a), we can find that the effect of the latter case is
absolutely superior to the first case. The advantages of the
part-offset subsidy policy can be explained as follows: when those
vaccinated individuals are offset by a certain proportion of
subsidy, from the individual's perspective, the cost of vaccination is
reduced. Thus more individuals are likely to take vaccination by
weighing the cost of vaccination and the cost of infection, and as
a result the density of infection is effectively inhibited.

\begin{figure}
\begin{center}
\includegraphics[width=6in]{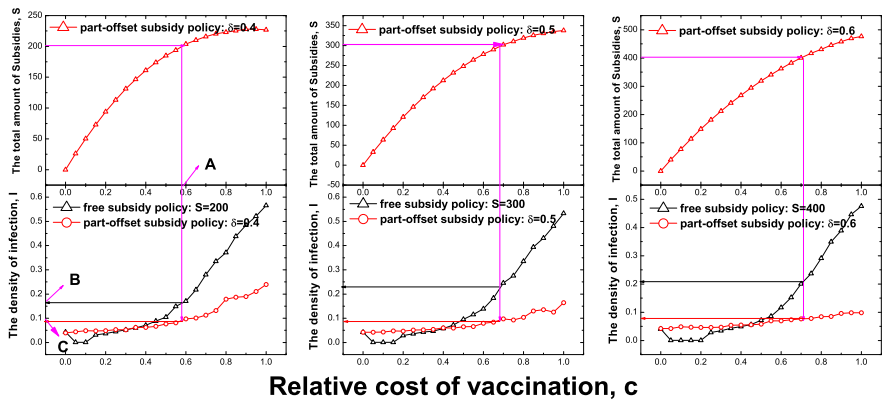}
\caption{
 Comparing the effects of the two subsidy
policies on the density of infection when the same amount of
subsidy is supplied: Left: $\delta=0.4$; center: $\delta=0.5$;
right: $\delta=0.6$. Top panel: the amount of subsidy as function
of $c$. Bottom panel: the density of infection as function of $c$.
}
\label{fig4}
\end{center}
\end{figure}

The effects of the two policies on controlling diseases under the same amount of subsidy are studied in figure~\ref{fig4}. Since the
total amount of subsidies for the part-offset subsidy policy is
determined by the number of vaccinated individuals and the value
of $\delta$, to fairly compare the two subsidy policies we first
preset a value of $S$ (e.g. $S=200$ in the left row of
figure~\ref{fig4}) and determine the corresponding $c$ (e.g. A in
the left row of figure~\ref{fig4}) in the part-offset subsidy
policy. Then the density of infection for the free subsidy policy
(e.g. B in the left row of figure~\ref{fig4} ) and the part-offset
subsidy policy (e.g. C in the left row of figure~\ref{fig4} ) at the
determined $c$ can be obtained respectively with the same amount
of subsidy. By choosing several representative values of $S$ in
figure~\ref{fig4}, we can find that the performance of the
part-offset subsidy policy is generally better than the free
subsidy policy. Moreover, the advantage of the part-offset
subsidy policy is more noticeable when $c$ is larger. We also
compare the efficiency of the two subsidy policies for other
values of $S$, and find that such phenomenon is universal. That is,
the effectiveness of the part-offset subsidy policy on controlling
the density of infection dominates.

\begin{figure}
\begin{center}
\includegraphics[width=4in]{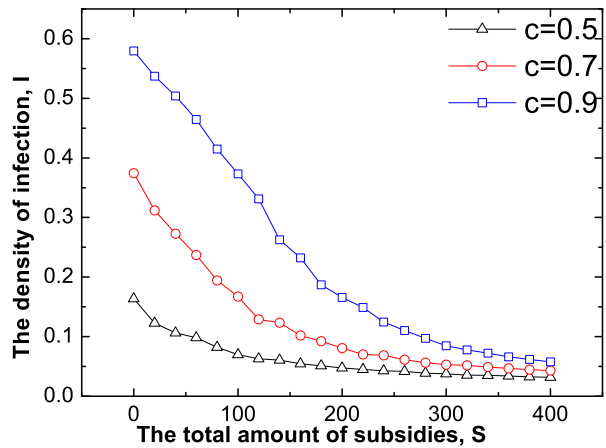}
\caption{
 For targeted free policy, the density of
infection as a function of $c$ for different values of
$S$. The donees are selected in a descending degree order, i.e., the nodes with higher degrees are vaccinated preferentially.}
\label{fig5}
\end{center}
\end{figure}

For the free subsidy policy, the beneficiaries are randomly
selected (labeled random free policy). Previous studies have
pointed out that the structure of contact network plays key roles
in the dynamics of an epidemic. For example, due to the existence
of hub nodes (nodes with very large degrees), epidemic can easily
spread on scale-free networks. Nonetheless, the structural
properties of a network can also give instructive hints toward
effective control of an epidemic. Targeted immunization was
considered by researchers and found that such
strategy can effectively prevent the outbreaks of diseases on
scale-free networks~\cite{pastor2003epidemics}. Here the targeted subsidy policy is also
considered (labeled targeted free subsidy policy), that is, the
individuals are freely vaccinated in a descending degree order
until the total amount of subsidy is used up. The density of
infection as a function of $S$ for different values of
$c$ is presented in figure~\ref{fig5}. As shown in figure~\ref{fig5}, with the
increase of $S$, the density of infection reduces quickly, and the
larger value of $c$, the better performance of targeted free subsidy
policy on reducing the density of infection.

\begin{figure}
\begin{center}
\includegraphics[width=6in]{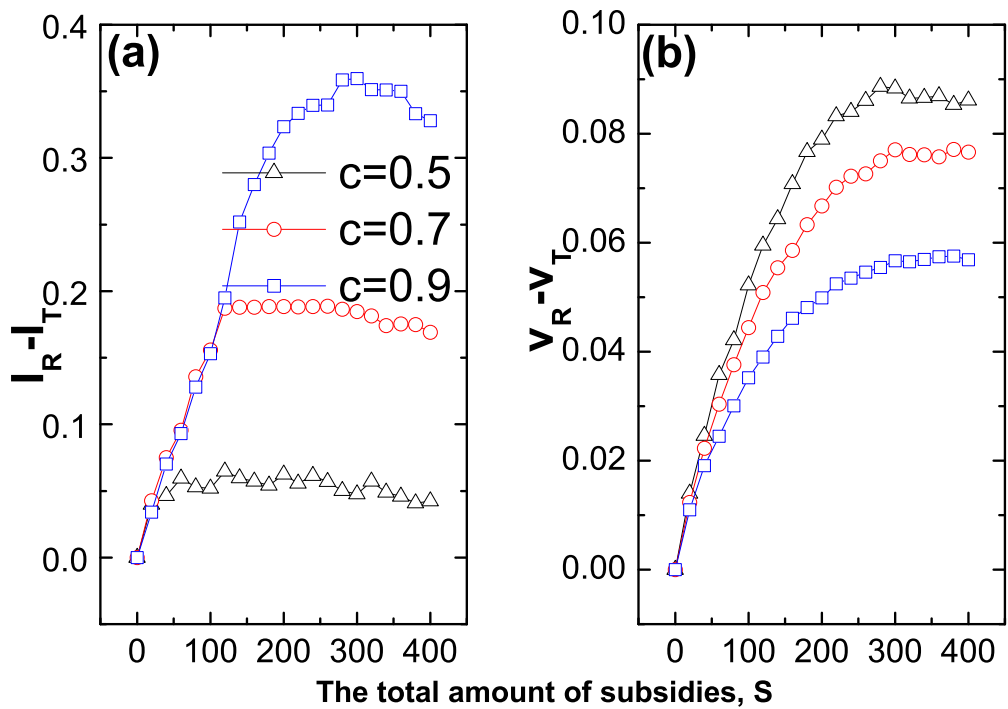}
\caption{
 The effects of the random free policy
and the targeted free policy are compared:
 (a) the difference
$I_R-I_T$ as a function of $S$. $I_R$ and $I_T$ are the density of
infection for random free policy and the targeted free policy
respectively; (b) the difference $V_R-V_T$ as a function of $S$. $V_R$ and $V_T$ are the fraction of vaccination for random free policy and targeted free policy respectively.}
\label{fig6}
\end{center}
\end{figure}

The efficiency of random and targeted free subsidy policies are
compared in figure~\ref{fig6} for different values of $c$.
Figure~\ref{fig6}(a) plots the difference of the density of
infection between the random and the targeted free policy
($I_R-I_T$), which indicates that the targeted free subsidy policy
is much better than the corresponding random policy in suppressing
the epidemic spreading when $S$ is large. Notably, the larger value
of $c$, the bigger the difference between the two approaches.

One might think that the number of vaccinated individuals in the
case of targeted free subsidy policy should be larger than that in the
random free subsidy policy. However, Figure~\ref{fig6}(b) shows
that the fraction of vaccination for the targeted free policy is
less than the case of random free policy (i.e. $V_R-V_T>0$). Such
phenomenon can be explained as follows: on one hand, when the
random policy is implemented, the individuals with large
degrees cannot be preferentially vaccinated. Thus, according to
Eq.~(\ref{eq6}), these `large-degree individuals' among non-donees
are likely to voluntarily choose to take vaccination; on the other hand, for the case of
targeted policy, the `large-degree individuals' have been
freely vaccinated, and only the individuals with smaller degrees
are not vaccinated. These `small-degree individuals' among
non-donees are less likely to take vaccination owing to the lower
perceived risk of infection. Thus, there exist more vaccinated
individuals in the random subsidy policy. Nonetheless, since the
`large-degree individuals' are freely vaccinated in the targeted
free subsidy policy when $S$ is rather large, the effectiveness of targeted
free subsidy policy is superior to the random one, irrespective of the
less number of vaccinated individuals.

From the perspective of group interest, the purpose of subsidy
policies is to minimize the total cost from both vaccination and
infection or the prevalence of disease. In many cases, the total
amount of the available subsidies is limited. In this case, how to
maximize the utility of the limited subsidies is an important and
significant task. For part-offset subsidy policy, we will
investigate what proportion $\delta$ of allowance can produce the
best possible results if the total amount of subsidies is limited.

\begin{figure}
\begin{center}
\includegraphics[width=6in]{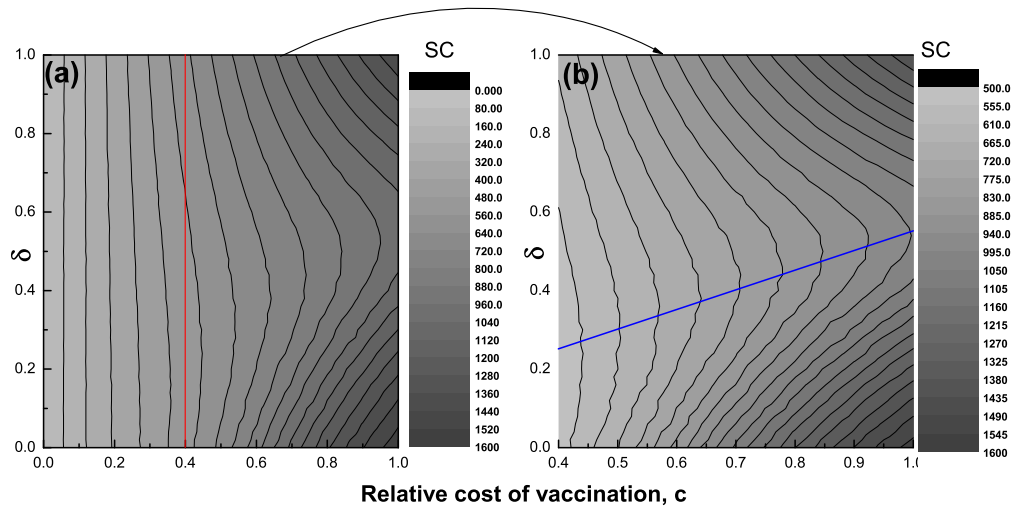}
\caption{
For part-offset subsidy policy, the total
social cost $SC$ as a function of the relative cost of vaccination $c$
and the proportion subsidy $\delta$: (a) $c$ from 0 to 1; (b) $c$
from 0.4 to 1. In (b) we redraw the result of (a) from $c=0.4$ (the right
side of red line in (a)).
The blue line given in (b) is to show the optimal $\delta$ for each value of $c$.}
\label{fig7}
\end{center}
\end{figure}

According to Eqs~(\ref{eq3}) and (\ref{eq4}), the cost of vaccination relative to infection is $c$, thus
the social cost from the group level can be defined
as~\cite{Bauch04,dybiec2004controlling}:
\begin{equation}\label{7}
SC=N_R\times1+((1-\delta)c+c\delta)N_V=N_R+N_Vc,
\end{equation}
where $N_R$ and $N_V$  are the number of recovered and vaccinated
persons. The impacts of $c$ and $\delta$ on the social cost $SC$ are
presented in figure~\ref{fig7}. From figure~\ref{fig7}(a), one can
observe that the value of $\delta$ has no significant effect on
the social cost $SC$ when $c$ is small. It is because the epidemic
can be controlled by voluntary vaccination when $c$ is small. As a
result, the effect of subsidy is negligible. When $c$ is large
(the left of red line in figure~\ref{fig7}(a)), we can find that
there exist an optimal value of $\delta$ leading to the lowest level
of $SC$ for every value of $c$. To explicitly reveal the
relationship between the optimal value of $\delta$ and the value
of $c$, Figure~\ref{fig7}(b) redraws the result of
figure~\ref{fig7}(a) starting from $c=0.4$. We can find that the
the optimal value of $\delta$ increase monotonically with $c$ (see
the blue line in figure~\ref{fig7}(b)). That is, the larger value of
$c$, the more subsidy proportion should be offset. We want to
stress that, for rather large value of $c$, the optimal value of
$\delta$ is neither 0 nor 1. In other words, 0\% or 100\% subsidy
policy does not minimize the total social cost.

\begin{figure}
\begin{center}
\includegraphics[width=6in]{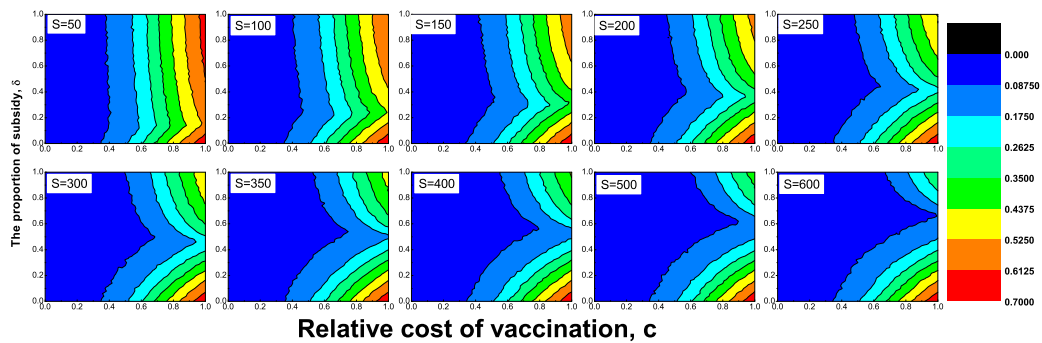}
\caption{
 For part-offset subsidy policy, the
density of infection as a function of the relative cost of
vaccination $c$ and the proportion subsidy $\delta$ for different
values of $S$. For fixed value of $S$, if the actual demanded
subsidy is larger than $S$, some vaccinated individuals are randomly
subsidized with proportion $\delta$ until $S$ is exhausted, and other
vaccinated individuals are not subsidized any more.}
\label{fig8}
\end{center}
\end{figure}

For some fatal diseases, the ultimate goal of the government is to
minimize the density of infection. If the total amount of
subsidies $S$ is fixed, what value of $\delta$ is able to control
the prevalence of disease to the lowest level? Figure~\ref{fig8}
shows that there exist an optimal values of $\delta$ leading to
the lowest level of infection for every value of $S$. Moreover,
the optimal value of $\delta$ increases with $S$. We can explain
such a phenomenon from two aspects: on one hand, when zero or very
small proportion of subsidy is offered,  the total amount of
subsidies would not be completely exhausted. In such a situation,
the subsidy policy can not exert significant impact on the
epidemic spreading; on the other hand, if $\delta$ is very large,
there will be only a few of individuals to get subsidized, similar
to the free subsidy policy, which failed to encourage more
individuals to take vaccination. Using the limited subsidy to
subsidize as many individuals as possible can guide more people to
take vaccination. Thus, we can find that, the small value of $S$,
the lower optimal value of $\delta$, which guarantees more
individuals to be offset.

\section{Conclusions and Discussions}\label{sec:discussion}

Incorporating behavioral response into epidemiological disease
spreading models can enhance a model's utility in evaluating
control methods. Models based on evolutionary game theory provide
a promising approach to investigate the complex interaction
between human behavior and the disease transmission process. In
this work, a network-based model was proposed to address the
impact of different subsidy policies on vaccination coverage and
disease prevalence when a voluntary vaccination mechanism is
considered. The results reveal that, without any external
incentives, diseases will prevail in the contact networks. Two
different subsidy policies---free subsidy policy and part-offset
subsidy policy are studied in detail. We found that the free
subsidy policy is not able to encourage more non-donees to take
vaccination, owing to the `herd immunity' coming from those freely
subsidized individuals. Thus, the performance of such a policy on
suppressing disease prevalence is not remarkable. One possible way
of reducing the prevalence of disease for this subsidy policy is
to subsidize individuals in a descending degree order, that is,
the individuals with more neighbors have higher probability of
being subsidized. Of course, this preferential subsidy policy require global information of the contact network, i.e., the
number of neighbors for each individual. In reality, this is
unlikely to be feasible --- but such a strategy could be
approximated by identifying ``high-risk'' groups. In contrast, we
have shown that: the effect of the part-offset subsidy policy is
prominent, it can effectively promote the coverage of vaccination
with fewer subsidies; and, it evidently reduces the prevalence of
disease. For the part-offset subsidy policy we also studied how to
maximize the effect of subsidy policy when the total amount of
subsidies is limited. Both from the total social cost and the
density of infection in population, our results suggested that the
best method is to distribute the limited resource to as many
individuals as possible.

How individuals deal with each new round of infectious disease is
often influenced by many external factors, such as rumors, mass
media, online information, and publishing the government information in a timely
manner~\cite{funk10,Perra11,funk09,ee}.
 In this work, we take a first step towards understanding
the impacts of different subsidy policies on vaccination coverage
and epidemiological dynamics. Some simplifying assumptions have
been made in the model. For example, our work assumes that the
individuals act according to self-interest and make decisions that
maximize their personal payoffs. In addition, we also assume that
the individuals have perfect information about the transmission
rate and the number of non-vaccinated individuals. All of these
assumptions were made in order to be able to keep the analysis
manageable, and to be able to focus in detail on the roles of
different subsidies.

To refine our understanding of vaccination behavior dynamics and
the roles of different subsidy polices, further work could be
developed in several aspects. First, how important effects of
subsidy policies on the control achievement are affected when some
individuals update strategies based on the payoff maximization
rule and others follow a different rule, such as imitation rule,
majority rule or minority rule, and so on. Second, given that some degree of
altruistic behavior is a common phenomenon in our society, we should ask how
the roles of the subsidy policy are modified when the altruism is taken
into account~\cite{Galvani12a}. Third, the effects of the rumors
and fears on the vaccination behavior dynamics are also worth
consideration.

The disease prevalence, the vaccinating behavior of individuals,
and the regulatory measures from the government are all interdependent
with each other, and hence should not be studied independently. As our
findings indicate, before implementing certain regulatory
measures, the government should also consider the effects of the
human behavioral responses. If an inappropriate subsidy policy is
adopted by the government, then the policy can not achieve the
expected results and may generate unnecessary waste.

\section*{Acknowledge}
This research was funded in part by the National Natural Science
Foundation of China (Grant Nos. 11005001, 11005051, 61004104,
61104143, 61004101 and 10975126), The Hong Kong Research Grants Council
Competitive Earmarked Research Grants(B-Q28N), and the 211 Project
of Anhui University(2009QN003A, KJTD002B).

\section*{References}

\bibliography{reference}

\bibliographystyle{iopart-num}

\end{document}